\documentclass[11pt,epsf]{article}

\usepackage{epsfig}
\usepackage{amssymb}
\usepackage{graphicx}
\usepackage{color}
\usepackage{subfigure}
\usepackage{mathtools}

\makeatletter

\@addtoreset{equation}{section} \makeatother
\def\b{\beta}
\def\g{\gamma}

\def\D{\Delta}

\def\Ph{\Phi}

\def\l{\lambda}
\def\L{\Lambda}
\def\m{\mu}
\def\n{\nu}

\def\s{\sigma}

\def\ta{\tau}

\def\pr{\prime}

\def\w{\wedge}

\def\lt{\left}
\def\rt{\right}

\def\nn{\nonumber}

\DeclarePairedDelimiter\abs{\lvert}{\rvert}%

\begin{document}

\begin{titlepage}
\title{\vskip -60pt
\vskip 20pt $T^3$ deformations and $\b$-deformed geometries }
\author{
Sunyoung Shin\footnote{e-mail : shin@theor.jinr.ru}}
\date{}
\maketitle \vspace{-1.0cm}
\begin{center}
~~~
\it Bogoliubov Laboratory of Theoretical Physics, JINR, 141980
Dubna, Moscow region, Russia
~~~\\
~~~\\
\end{center}

\begin{abstract}
We discuss $\b$-deformed geometries on two types of $T^3$'s where
the direction along the third coordinate is not orthogonal to the
direction along the second coordinate or the direction along the
first coordinate. We show that the intersection angle between the
direction along the third coordinate and the direction along the
second coordinate corresponds to the parameter of the S-duality of
the $\b$-deformation while the intersection angle between the
direction along the third coordinate and the direction along the
first coordinate generalizes the $\b$-deformed geometry.
\end{abstract}

\end{titlepage}
\newpage


\section{Introduction}
\setcounter{equation}{0}
Type IIB supergravity on a two-torus has an
$E_{3,3}=SL(3,\mathbb{R})\times SL(2,\mathbb{R})$ symmetry in the
classical low-energy limit. It is shown that a specific
$SL(3,\mathbb{R})$ transformation generates the gravity dual
\cite{Lunin:2005jy} of the marginal deformation \cite{Leigh:1995ep}.
The $SL(3,\mathbb{R})$ is geometric from the eleven dimensional
viewpoint. In type IIB supergravity, it consists of an
$SL(2,\mathbb{R})$ acting on the K\"{a}hler structure modulus and an
S-duality transformation $SL(2,\mathbb{R})_s$. A particular
$SL(2,\mathbb{R})$ transformation on a two-torus produces a
non-singular geometry if the original geometry is non-singular,
which is the gravity description of the $\gamma$-deformation of the
gauge theory to introduce phases in the superpotential. This
corresponds to a TsT (T-duality, shift, T-duality) transformation.
It breaks the supersymmetries which depend on the coordinates of the
two-torus. The $\beta$-deformation of the gauge theory is obtained
by complexifying the parameter in the superpotential. In the gravity
side, the deformation corresponds to a particular $SL(3,\mathbb{R})$
transformation, which is formed by the $SL(2,\mathbb{R})$ and an
S-duality transformation $SL(2,\mathbb{R})_s$ or equivalently STsTS
(S-duality, T-duality, shift, T-duality, S-duality). Since the
$SL(2,\mathbb{R})$ is the nongeometric part of
$SO(2,2,\mathbb{R})\simeq SL(2,\mathbb{R})\times SL(2,\mathbb{R})$,
which is a subgroup of the type IIB T-duality group, the
$SL(2,\mathbb{R})$ can be viewed as $O(2,2,\mathbb{R})$ acting on
the background matrix $E=g+B$ \cite{Giveon:1994fu}. The
corresponding $O(2,2,\mathbb{R})$ matrices for the deformations
discussed in \cite{Lunin:2005jy,Frolov:2005dj} are identified in
\cite{CatalOzer:2005mr}. Multiparameter deformations of them are
discussed in \cite{Frolov:2005dj,Fokken:2013aea}. General formulae
for TsT transformations and examples of the deformed gauge theories
and their gravity duals are discussed in \cite{Imeroni:2008cr}.
Various aspects of marginal deformations of gauge theories and the
gravity duals of them are investigated in
\cite{Aharony:2002tp,Frolov:2005ty,Ahn:2005vc,Gauntlett:2005jb,deMelloKoch:2005vq,
Pal:2005nr,Chu:2006tp,Berman:2007tf,Ahn:2012hsa,Ahn:2012hs}.

In \cite{Lunin:2005jy}, the type IIB supergravity solution is
obtained from eleven dimensional supergravity on a rectangular
three-torus ($\varphi^1$,$\varphi^2$,$\varphi^3$) by a dimensional
reduction along $\varphi^3$ and a T-duality transformation along
$\varphi^1$. The new geometry \cite{Lunin:2005jy} is generated by
the $SL(2,\mathbb{R})$ or $SL(3,\mathbb{R})$ transformation of the
type IIB supergravity solution.

We study the $SL(2,\mathbb{R})$ or $SL(3,\mathbb{R})$ transformation
on deformed $T^3$'s. We deform a rectangular three-torus by two
types of $SL(3,\mathbb{R})$ transformations so that the direction
along the third coordinate is not orthogonal to the direction along
the second coordinate or the direction along the first coordinate.
By a dimensional reduction along the direction along the third
coordinate and a T-duality transformation along the direction along
the first coordinate after the transformations, we obtain geometries
with the intersection angles of the directions as parameters. We
discuss the role of the parameters under the $SL(2,\mathbb{R})$ or
$SL(3,\mathbb{R})$ transformation which generates the $\b$-deformed
geometry \cite{Lunin:2005jy}.

In section \ref{sec:torus_def}, we review Lunin and Maldacena's
solution generating technique and discuss the two types of
$SL(3,\mathbb{R})$ transformations. In section
\ref{sec:torus_def23}, we study the type IIB supergravity solution
obtained by a dimensional reduction along the direction along the
third coordinate, which is not orthogonal to the direction along the
second coordinate, followed by a T-duality transformation along the
direction along the first coordinate. We show that the intersection
angle forms the S-duality transformation of \cite{Lunin:2005jy}
under the TsT transformation. In section \ref{sec:torus_def13}, we
study the type IIB supergravity solution obtained by a dimensional
reduction along the direction along the third coordinate, which is
not orthogonal to the direction along the first coordinate, followed
by a T-duality transformation along the direction along the first
coordinate. We show that the intersection angle generalizes the
$\b$-deformed geometry. In section \ref{sec:discuss}, we summarize
our results.

%
\section{$T^3$ deformation}\label{sec:torus_def}
\setcounter{equation}{0}
The type IIB supergravity solution with $U(1)\times U(1)$ symmetry
\cite{Lunin:2005jy} is derived from an eleven-dimensional
supergravity solution with $U(1)\times U(1)\times U(1)$ symmetry .
The coordinates of the three-torus are
$(\varphi^1,\varphi^2,\varphi^3)$. The type IIB supergravity
solution is obtained by a dimensional reduction along $\varphi^3$
and a T-duality transformation along $\varphi^1$
\cite{Lunin:2005jy}:
\begin{eqnarray}
ds^2_{IIB}&=&F\lt[\frac{1}{\sqrt{\D}}\lt(D\varphi^1-CD\varphi^2\rt)^2
+\sqrt{\D}(D\varphi^2)^2\rt]+\frac{\ta_2^{-2/3}}{F^{1/3}}g_{\m\n}dx^\m dx^\n, \nn\\
B^{(2)}&=&B_{12}D\varphi^1 \w D\varphi^2+\lt(B_{1\m}
D\varphi^1+B_{2\m} D\varphi^2\rt) \w dx^\mu-\frac{1}{2}\lt(A^m_\m
B_{m\n}-\tilde{b}_{\m\n}\rt) dx^\m \w dx^\n, \nn\\
C^{(0)}&=&\ta_1, \nn\\
C^{(2)}&=&C_{12}D\varphi^1 \w
D\varphi^2+\lt(C_{1\m}D\varphi^1+C_{2\m}D\varphi^2\rt) \w dx^\m
-\frac{1}{2}\lt(A^m_\m C_{m\n}-\tilde{c}_{\m\n}\rt)dx^\m\w
dx^\n,  \nn\\
C^{(4)}&=&-\frac{1}{2}\lt(\tilde{d}_\m+B_{12}\tilde{c}_{\m\n}
-\epsilon^{mn}B_{m\m}C_{n\n}-B_{12}A^m_\m
C_{m\n}\rt)dx^\m \w dx^\n \w D\varphi^1 \w D\varphi^2\nn\\
&&+\frac{1}{6}\lt[C_{\m\n\l}+3\lt(\tilde{b}_{\m\n}+A_\m^1B_{1\n}
-A_\m^2B_{2\n}\rt)C_{1\l}\rt]dx^\m\w
dx^\n \w dx^\l \w D\varphi^1\nn\\
&&+d_{\m_1\m_2\mu_3\mu_4}dx^{\m_1} \w dx^{\m_2} \w dx^{\m_3} \w
dx^{\m_4} + \hat{d}_{\m_1 \m_2 \m_3} dx^{\m_1} \w dx^{\m_2} \w
dx^{\m_3} \w D\varphi^2,  \nn
\end{eqnarray}
\begin{eqnarray} \label{eq:IIBa}
D\varphi^1\equiv d\varphi^1+A_\m^1 dx^\m,~~~D\varphi^2\equiv
d\varphi^2+A_\m^2 dx^\mu,
\end{eqnarray}
\begin{eqnarray}\label{eq:IIBb}
&&M=gg^T,\nn\\
&&g^T=\lt(
\begin{array}{ccc}
\ta_2^{1/3}F^{-1/3}    &           0            &   0    \\
0                     &  \ta_2^{1/3}F^{2/3}     &   0    \\
0                     &           0            &
\ta_2^{-2/3}F^{-1/3}
\end{array}\rt)\lt(
\begin{array}{ccc}
1       &       B_{12}      &            0\\
0       &       1           &            0 \\
\ta_1    & -C_{12}+\ta_1 B_{12}   &   1
\end{array}\rt).
\end{eqnarray}
Under an $SL(3,\mathbb{R})$ transformation
\begin{eqnarray}
\lt(\begin{array}{c} %
\varphi^1 \\
\varphi^2 \\
\varphi^3 \\
\end{array}\rt)^\pr=%
\lt(\L^T\rt)^{-1}
\lt(\begin{array}{c} %
\varphi^1 \\
\varphi^2 \\
\varphi^3 \\
\end{array}\rt),
\end{eqnarray}
the metric transforms as
\begin{eqnarray}
M\rightarrow \L M \L^T=g^\pr g^{\pr \, T},
\end{eqnarray}
\begin{eqnarray} \label{eq:transf_m}
g^{\pr \, T}=\lt(%
\begin{array}{ccc}
\frac{1}{\sqrt{G H}}\ta_2^{1/3}F^{-1/3}   &       0      &    0   \\
0     &  \sqrt{G}\ta_2^{1/3}F^{2/3}                 &                 0           \\
0     &               0                &  \sqrt{H}
\ta_2^{-2/3}F^{-1/3}
\end{array}\rt)%
\lt(\begin{array}{ccc}
1                 &          B^\pr_{12}           &        0            \\
0                 &             1                           &        0            \\
\chi^\pr   &      -C^\pr_{12}+\chi^\pr B^\pr_{12}        &  1
\end{array}\rt),
\end{eqnarray}
and the fields
\begin{eqnarray}
V_\m^{(1)}=\lt( %
\begin{array}{c}
-B_{2\m}  \\
A_\m^1    \\
C_{2\mu}
\end{array}
\rt),~~V_\m^{(2)}=\lt(
\begin{array}{c}
B_{1\mu}   \\
A_\m^2     \\
-C_{1\mu}
\end{array}\rt),~~
W_{\m\n}=\lt(
\begin{array}{c}
\tilde{c}_{\m\n}   \\
\tilde{d}_{\m\n}   \\
\tilde{b}_{\m\n}
\end{array}\rt),
\end{eqnarray}
transform as vectors
\begin{eqnarray} \label{eq:transf_f}
V_\m^{(i)}\rightarrow (\L^T)^{-1}V_\m^{(i)},~~W_{\m\n}\rightarrow \L
W_{\m\n}.
\end{eqnarray}
The scalars $\D$, $C$ and the three form $C_{\m\n\l}$ stay
invariant. The $SL(3,\mathbb{R})$ matrix \cite{Lunin:2005jy}, which
generates the gravity duals of the $\b$-deformations is
\begin{eqnarray} \label{eq:LM_matrix}
\L_{LM}^T=\lt(%
\begin{array}{ccc}
1      &    0    &    0   \\
\g     &    1    &    \s  \\
0      &    0    &    1
\end{array}
\rt).
\end{eqnarray}

We investigate two types of type IIB supergravity solutions. One is
a type IIB supergravity solution which is obtained from an eleven
dimensional supergravity solution living on a slanted three-torus
where the direction along the third coordinate is not orthogonal to
the direction along the second coordinate. The other is a type IIB
supergravity solution which is obtained from an eleven dimensional
supergravity solution living on a slanted three-torus where the
direction along the third coordinate is not orthogonal to the
direction along the first coordinate. The torus deformation can be
done by an $SL(3,\mathbb{R})$ transformation
\begin{eqnarray}
\lt(\begin{array}{c} %
\varphi^1 \\
\varphi^2 \\
\varphi^3 \\
\end{array}\rt)^\pr=%
\lt(L\rt)^{-1}
\lt(\begin{array}{c} %
\varphi^1 \\
\varphi^2 \\
\varphi^3 \\
\end{array}\rt),~~~L=L_1,~L_2\in SL(3,\mathbb{R}),
\end{eqnarray}
with
\begin{eqnarray}\label{eq:l1l2}
L_1=\lt(%
\begin{array}{ccc}
1     &    0   &  0    \\
0     &    1   &  (r_3/R_2)\cos\xi \\
0     &    0   &  1
\end{array}
\rt),~~~
L_2=\lt(%
\begin{array}{ccc}
1     &    0   &  (r_3/R_1)\cos\xi    \\
0     &    1   &  0 \\
0     &    0   &  1
\end{array}
\rt),
\end{eqnarray}
which are constrained by
\begin{eqnarray}
r_3=\frac{R_3}{\sin\xi}.
\end{eqnarray}
$R_i,(i=1,2,3)$ are the radii of the $T^3$ before the
transformations and $r_3$ is the radius of the third direction after
the transformations. We study the type IIB supergravity solutions
produced by a dimensional reduction along the direction along the
third coordinate and a T-duality transformation along the direction
along the first coordinate by using (\ref{eq:transf_m}) and
(\ref{eq:transf_f}). We assume that only the metric, complex field
$\ta=\ta_1+i\ta_2$ and $\tilde{d}_{\m\n}$ are excited and the other
fields are zero.

\section{$T^3$ deformed by $L_1$}\label{sec:torus_def23} \setcounter{equation}{0}
We deform a rectangular three-torus by $L_1$ in (\ref{eq:l1l2})
\begin{eqnarray} \label{eq:sll1}
L_1=\lt(%
\begin{array}{ccc}
1     &    0   &  0    \\
0     &    1   &  L^2_{~3} \\
0     &    0   &  1
\end{array}
\rt),%
\end{eqnarray}
where
\begin{eqnarray}
L^2_{~3}=\frac{r_3}{R_2}\cos\xi=k^{-1}\cot\xi,~~k=\frac{R_2}{R_3}.
\end{eqnarray}
For fixed $R_2$ and $\xi$, the torus radius $r_3$ can be chosen so
that $L^2_{~3}\sim L^2_{~3}+1$.

By using (\ref{eq:transf_m}) and (\ref{eq:transf_f}), we find the
geometry in ten dimensions with the intersection angle $\xi$ as a
parameter:
\begin{eqnarray}
\bar{G}^{-1}&=&1+\lt(\ta_1^2+\ta_2^2\rt) k^{-2}\cot^2\xi  F^2, \nn\\
H&=&1+\ta_2^2 k^{-2}\cot^2\xi F^2, \nn
\end{eqnarray}
\begin{eqnarray}\label{eq:IIB23a}
d\bar{s}^2&=&\bar{F}\lt[\frac{1}{\sqrt{\D}}\lt(D\varphi^1-CD\varphi^2\rt)^2
+\sqrt{\D}\lt(D\varphi^2\rt)^2\rt]
+\frac{e^{2\bar{\Ph}/3}}{\bar{F}^{1/3}}g_{\m\n}dx^\m dx^\n,   \nn\\
\bar{F}&=&F\bar{G}\sqrt{H}, \nn\\
e^{\bar{\Ph}}&=&\sqrt{\bar{G}}H\ta_2^{-1}, \nn\\
\bar{\chi}&=&H^{-1}\ta_1, \nn\\
\bar{B}^{(2)}&=&- \bar{G} F^2\ta_1 k^{-1}\cot\xi D\varphi^1 \w
D\varphi^2 +\frac{1}{2}\tilde{d}_{\m\n} k^{-1}\cot\xi
dx^\m \w dx^\n,\nn\\
\bar{C}^{(2)}&=&- \bar{G} F^2\lt(\ta_1^2+\ta_2^2\rt) k^{-1}\cot\xi
D\varphi^1 \w D\varphi^2, \nn\\
\bar{C}^{(4)}&=&-\frac{1}{2}\tilde{d}_{\m\n}D\varphi^1 \w D\varphi^2
\w dx^\m \w dx^\n.
\end{eqnarray}

We show that the intersection angle $\xi$ forms the representation
of the S-duality $SL(2,\mathbb{R})_s$ with the parameter $\g$ of the
$SL(2,\mathbb{R})$ transformation considered in \cite{Lunin:2005jy}.
It can be guessed since the matrix (\ref{eq:LM_matrix}) is
factorized as
\begin{eqnarray}
\L_{LM}^T=\lt(%
\begin{array}{ccc}
1      &    0    &    0   \\
\g     &    1    &    \s  \\
0      &    0    &    1
\end{array}
\rt)=
\lt(%
\begin{array}{ccc}
1      &    0    &    0   \\
0     &    1    &    \s  \\
0      &    0    &    1
\end{array}
\rt)
\lt(%
\begin{array}{ccc}
1      &    0    &    0   \\
\g      &    1    &   0  \\
0      &    0    &    1
\end{array}
\rt).
\end{eqnarray}
The component $L^2_{~3}$ of (\ref{eq:sll1}) corresponds to $\s$. We
discuss the $\b$-deformed geometry. The $SL(2,\mathbb{R})$
transformation can be realized as the TsT transformation
\cite{Lunin:2005jy,Frolov:2005dj} or $O(2,2,\mathbb{R})$
\cite{Lunin:2005jy,CatalOzer:2005mr} acting on the background
matrix. We present the TsT transformation. By a T-duality
transformation along $\varphi^1$, a shift transformation
$\varphi^2\rightarrow \varphi^2+\g \varphi^1$ and a T-duality
transformation along $\varphi^1$, we obtain a geometry
\begin{eqnarray} \label{eq:IIB23b}
d\bar{s}^{\pr 2} &=&
\frac{\bar{F}}{(1+\g\bar{B}_{12})^2+\g^2\bar{F}^2}
\lt[\frac{1}{\sqrt{\D}}\lt(D\varphi^1-CD\varphi^2\rt)^2+\sqrt{\D}\lt(D\varphi^2\rt)^2
\rt]
+\frac{e^{2\bar{\Phi}/3}}{\bar{F}^{1/3}}g_{\m\n}dx^\m dx^\n,   \nn\\
e^{2\bar{\Ph}^\pr}&=&\frac{e^{2\bar{\Ph}}}{\lt(1+\g \bar{B}_{12}\rt)^2+\g^2\bar{F}^2}, \nn\\
\bar{B}^{(2)\pr} &=&
\frac{\bar{B}_{12}+\g\lt(\bar{B}_{12}^2+\bar{F}^2\rt)}{\lt(1+\g\bar{B}_{12}\rt)^2+\g^2\bar{F}^2}
D\varphi^1 \w D\varphi^2+\frac{1}{2}\tilde{d}_{\mu\nu}k^{-1}\cot\xi
dx^\m \w
dx^\n, \nn\\
\bar{\chi}^\pr&=& \bar{\chi}-\g\lt(\bar{C}_{12}-\bar{\chi}\bar{B}_{12}\rt), \nn\\
\bar{C}^{(2)\pr}&=&\frac{\bar{C}_{12}(1+\g\bar{B}_{12})+\g\bar{\chi}\bar{F}^2}{\lt(1+\g\bar{B}_{12}\rt)^2+\g^2\bar{F}^2}
D\varphi^1 \w D\varphi^2
+\frac{1}{2}\g\tilde{d}_{\m\n}dx^\m \w dx^\n, \nn\\
\bar{C}^{(4)\pr} &=&
-\frac{1}{2}\tilde{d}_{\m\n}\frac{1+\g\bar{B}_{12}}{\lt(1+\g\bar{B}_{12}\rt)^2+\g^2\bar{F}^2}
D\varphi^1 \w D\varphi^2 \w dx^\m \w dx^\n.
\end{eqnarray}
By introducing
\begin{eqnarray}
G^{-1}&=&1+(\g-\ta_1 k^{-1}\cot\xi)^2F^2+\ta_2^2 k^{-2}\cot^2\xi
F^2,
\end{eqnarray}
and using the relations in (\ref{eq:IIB23a}), the geometry
(\ref{eq:IIB23b}) can be rewritten as
\begin{eqnarray} \label{eq:IIB23c}
d\bar{s}^{\pr 2} &=& \bar{F}^\pr
\lt[\frac{1}{\sqrt{\D}}\lt(D\varphi^1-CD\varphi^2\rt)^2+\sqrt{\D}\lt(D\varphi^2\rt)^2
\rt]
+\frac{e^{2\bar{\Ph}^\pr/3}}{\bar{F}^{\pr1/3}}g_{\m\n}dx^\m dx^\n,   \nn\\
\bar{F}^\pr &=& FG\sqrt{H},  \nn\\
e^{\bar{\Ph}^\pr} &=& \sqrt{G}H\ta_2^{-1}, \nn\\
\bar{B}^{(2)\pr}  &=& G F^2(\g-\ta_1k^{-1}\cot\xi)D\varphi^1 \w
D\varphi^2+\frac{1}{2}\tilde{d}_{\mu\nu}k^{-1}\cot\xi dx^\m \w
dx^\n,  \nn\\
\bar{\chi}^\pr &=&H^{-1}(\ta_1+\g\ta_2^2 k^{-1}\cot\xi F^2), \nn\\
\bar{C}^{(2)\pr} &=& G F^2\lt[\ta_1\g-(\ta_1^2+\ta_2^2)
k^{-1}\cot\xi \rt] D\varphi^1 \w D\varphi^2
+\frac{1}{2}\g\tilde{d}_{\m\n}dx^\m \w dx^\n \nn\\
%
F^{(5)}&=&\tilde{F}^{(5)}+\star\tilde{F}^{(5)},
\end{eqnarray}
where the star is taken with the new metric.

This is the $\beta$-deformed geometry \cite{Lunin:2005jy} with
$\s=k^{-1}\cot\xi$. For $\xi=\frac{\pi}{2}$, the geometry
(\ref{eq:IIB23c}) is the gravity dual of the $\g$-deformation.

\section{$T^3$ deformed by $L_2$}\label{sec:torus_def13} \setcounter{equation}{0}
We have shown that the intersection angle between the direction
along the third coordinate and the direction along the second
coordinate is the parameter for the S-duality of
(\ref{eq:LM_matrix}). It is therefore expected that the intersection
angle between the direction along the third coordinate and the
direction along the first coordinate is a parameter which
generalizes the $SL(3,\mathbb{R})$ transformation
(\ref{eq:LM_matrix}). We deform a rectangular three-torus by $L_2$
in (\ref{eq:l1l2})
\begin{eqnarray} \label{eq:sll2}
L _2=\lt(%
\begin{array}{ccc}
1   &   0   &  L^1_{~3} \\
0   &   1   &    0      \\
0   &   0   &    1
\end{array}
\rt),
\end{eqnarray}
where
\begin{eqnarray}
L^1_{~3}=\frac{r_3}{R_1}\cos\xi=l^{-1}\cot\xi,~~~l=\frac{R_1}{R_3}.
\end{eqnarray}
By using (\ref{eq:transf_m}) and (\ref{eq:transf_f}), we find the
geometry with the intersection angle $\xi$ as a parameter:

\begin{eqnarray}
\bar{G}^{-1}&=&1 \nn\\
\bar{H}&=&1+2l^{-1}\cot\xi\ta_1+l^{-2}\cot^2\xi
\abs{\ta}^2,~~~~~(\abs{\ta}^2=\ta_1^2+\ta_2^2), \nn
\end{eqnarray}
\begin{eqnarray}\label{eq:IIB13a}
d\bar{s}^2&=&\bar{F}\lt[\frac{1}{\sqrt{\D}}\lt(D\varphi^1-CD\varphi^2\rt)^2
+\sqrt{\D}\lt(D\varphi^2\rt)^2\rt]
+\frac{e^{2\bar{\Ph}/3}}{\bar{F}^{1/3}}g_{\m\n}dx^\m dx^\n,   \nn\\
\bar{F}&=&F\sqrt{\bar{H}}, \nn\\
e^{\bar{\Ph}}&=&\bar{H}\ta_2^{-1}, \nn\\
\bar{\chi}&=&\bar{H}^{-1}\lt(\ta_1+l^{-1}\cot\xi\abs{\ta}^2\rt), \nn\\
\bar{B}^{(2)}&=&0, \nn\\
\bar{C}^{(2)}&=&0, \nn\\
\bar{C}^{(4)}&=&-\frac{1}{2}\tilde{d}_{\m\n}D\varphi^1 \w D\varphi^2
\w dx^\m \w dx^\n.
\end{eqnarray}
The axion-dilaton field $\ta=\ta_1+i\ta_2$ transforms as
\begin{eqnarray} \label{eq:cplex_transf}
\bar{\ta}=\bar{\chi}+ie^{-\bar{\Ph}}=\frac{\ta}{l^{-1}\cot\xi\ta+1}.
\end{eqnarray}

The torus deformation provides a generalized $\b$-deformed geometry
under the $SL(3,\mathbb{R})$ transformation (\ref{eq:LM_matrix}) of
\cite{Lunin:2005jy}, since the matrix
\begin{eqnarray}
\L^T= \lt(%
\begin{array}{ccc}
1     &  0  &  l^{-1}\cot\xi  \\
\g    &  1  &  \s        \\
0     &  0  &  1
\end{array}
\rt),
\end{eqnarray}
is factorized as
\begin{eqnarray}
\L^T=L_2\,\L_{LM}^T.
\end{eqnarray}

The $SL(3,\mathbb{R})$ transformation (\ref{eq:LM_matrix}) of the
geometry (\ref{eq:IIB13a}) is
\begin{eqnarray}
G^{-1}&=&1+\lt(\g-\s\bar{\chi}\rt)^2\bar{F}^2+\s^2e^{-2\bar{\Ph}}\bar{F}^2, \nn\\
\bar{H}^\pr&=&1+\s^2e^{-2\bar{\Ph}}\bar{F}^2, \nn
\end{eqnarray}
\begin{eqnarray}\label{eq:IIB13b}
d\bar{s}^{\pr2}&=&\bar{F}^\pr\lt[\frac{1}{\sqrt{\D}}\lt(D\varphi^1-CD\varphi^2\rt)^2
+\sqrt{\D}\lt(D\varphi^2\rt)^2\rt]
+\frac{e^{2\bar{\Ph}^\pr/3}}{\bar{F}^{\pr1/3}}g_{\m\n}dx^\m dx^\n,   \nn\\
\bar{F}^\pr&=&\bar{F}G\sqrt{\bar{H}^\pr}, \nn\\
e^{\bar{\Ph}^\pr}&=&\sqrt{G}\bar{H}^\pr e^{\bar{\Ph}}, \nn\\
\bar{\chi}^\pr&=&\bar{H}^{\pr-1}\lt(\bar{\chi}+\g \s e^{-2\bar{\Ph}} \bar{F}^2\rt), \nn\\
\bar{B}^{\pr(2)}&=&G\bar{F}^2\lt(\g - \s \bar{\chi}\rt)D\varphi^1 \w
D\varphi^2
+\frac{\s}{2}\tilde{d}_{\m\n} dx^\m \w dx^\n, \nn\\
\bar{C}^{\pr(2)}&=&G\bar{F}^2\lt(\g \bar{\chi} - \s
\abs{\bar{\ta}}^2 \rt)D\varphi^1 \w D\varphi^2
+\frac{\g}{2}\tilde{d}_{\m\n} dx^\m \w dx^\n, \nn\\
F^{(5)}&=&\tilde{F}^{(5)}+\star\tilde{F}^{(5)}.
%
\end{eqnarray}
The star is taken with the new metric. By using (\ref{eq:IIB13a}),
the geometry (\ref{eq:IIB13b}) can be rewritten as
\begin{eqnarray}
G^{-1}&=&1+\lt(\g^2f-2\g\s h +\s^2g\rt)F^2, \nn\\
H&=&f+\ta_2^2\s^2F^2, \nn
\end{eqnarray}
\begin{eqnarray}\label{eq:IIB13c}
d\bar{s}^{\pr2}&=&\bar{F}^\pr\lt[\frac{1}{\sqrt{\D}}\lt(D\varphi^1-CD\varphi^2\rt)^2
+\sqrt{\D}\lt(D\varphi^2\rt)^2\rt]
+\frac{e^{2\bar{\Ph}^\pr/3}}{\bar{F}^{\pr1/3}}g_{\m\n}dx^\m dx^\n, \nn\\
\bar{F}^\pr&=&F G\sqrt{H},   \nn\\
e^{\bar{\Ph}^\pr}&=&\sqrt{G}H\ta_2^{-1},  \nn\\
\bar{\chi}^\pr&=&H^{-1}\lt(h+\g\s \ta_{2}^2F^2\rt), \nn\\
\bar{B}^{\pr(2)}&=&GF^2\lt(\g f -\s h\rt)D\varphi^1 \w D\varphi^2
+\frac{\s}{2}\tilde{d}_{\m\n}  dx^\m \w dx^\n,  \nn\\
\bar{C}^{\pr(2)}&=&GF^2\lt(\g h -\s g\rt)D\varphi^1 \w
D\varphi^2+\frac{\g}{2}\tilde{d}_{\m\n} dx^\m \w dx^\n,
\end{eqnarray}
where
\begin{eqnarray}\label{eq:IIB13fgha}
f&=&(1+l^{-1}\cot\xi\ta_1)^2+l^{-2}\cot^2\xi\ta_2^2,  \nn\\
g&=&\abs{\ta}^2,  \nn\\
h&=&\ta_{1}+l^{-1}\cot\xi\abs{\ta}^2.
\end{eqnarray}

The geometry (\ref{eq:IIB13c}) with $f$, $g$ and $h$ is valid for
any $SL(3,\mathbb{R})$ transformation of the form
\begin{eqnarray} \label{eq:sll2b}
&&L=\lt(%
\begin{array}{ccc}
L^1_{~1}   &   0   &  L^1_{~3} \\
0   &   1   &    0      \\
L^3_{~1}   &   0   &  L^3_{~3}
\end{array}
\rt),~~\det{L}=1,\nn\\
&&\L^T=L\,\L_{LM}^T=\lt(%
\begin{array}{ccc}
L^1_{~1}   &   0   &  L^1_{~3} \\
\g   &   1   &    \s      \\
L^3_{~1}   &   0   &  L^3_{~3}
\end{array}
\rt),
\end{eqnarray}
with
\begin{eqnarray}
f&=&\lt(L^3_{~3}+L^1_{~3}\ta_1\rt)^2+\lt(L^1_{~3}\rt)^2\ta_2^2, \nn\\
g&=&\lt(L^3_{~1}+L^1_{~1}\ta_1\rt)^2+\lt(L^1_{~1}\rt)^2\ta_2^2,   \nn\\
h&=&\lt(L^3_{~3}+L^1_{~3}\ta_1\rt)(L^3_{~1}+L^1_{~1}\ta_1)+L^1_{~1}L^1_{~3}\ta_2^2.
\end{eqnarray}

The generalization of the $\b$-deformed geometry corresponds to the
$SL(2,\mathbb{R})$ symmetry of type IIB theory as we have seen in
(\ref{eq:cplex_transf}). It is also the symmetry of the toroidal
compactification \cite{Schwarz:1995dk,Aspinwall:1995fw}. We consider
a case in which $\ta_1=0$ by using the shift symmetry of the type
IIB supergravity and $\ta_2=\frac{R_1}{R_3}$ by equating the
axion-dilaton field with the torus modulus of the rectangular torus
before the transformation as it is done in \cite{Schwarz:1995dk}.
Then $\ta=i\frac{R_1}{R_3}$ transforms under (\ref{eq:cplex_transf})
as
\begin{eqnarray} \label{eq:tau_mod}
\bar{\ta}=\frac{R_1}{r_3}e^{i\xi}.
\end{eqnarray}
This is the torus moduli of the two-torus $(\varphi^1,\varphi^3)$
deformed by (\ref{eq:sll2}) as expected.

We identify the complex scalar field $\ta$ with the torus modulus of
the rectangular torus before the deformation (\ref{eq:sll2}). Then
(\ref{eq:IIB13c}) becomes simpler. The geometry is
\begin{eqnarray}\label{eq:gen_grav_dual}
G^{-1}&=&1+\g^2 F^2+ \lt( \s l-\g\cot\xi \rt)^2F^2, \nn\\
H&=&\csc^2\xi+\s^2l^2F^2, \nn\\
d\bar{s}^{\pr2}&=&\bar{F}^\pr\lt[\frac{1}{\sqrt{\D}}\lt(D\varphi^1-CD\varphi^2\rt)^2
+\sqrt{\D}\lt(D\varphi^2\rt)^2\rt]
+\frac{e^{2\bar{\Ph}^\pr/3}}{\bar{F}^{\pr1/3}}g_{\m\n}dx^\m dx^\n, \nn\\
\bar{F}^\pr&=&F G\sqrt{H},   \nn\\
e^{\bar{\Ph}^\pr}&=&\sqrt{G}H l^{-1},  \nn\\
\bar{\chi}^\pr&=&\frac{1}{H}\lt(l\cot\xi+\g\s l^2F^2\rt), \nn\\
\bar{B}^{\pr(2)}&=&GF^2 \lt(\g \csc^2\xi-\s l\cot\xi\rt)D\varphi^1
\w D\varphi^2
+\frac{\s}{2}\tilde{d}_{\m\n}  dx^\m \w dx^\n,  \nn\\
\bar{C}^{\pr(2)}&=&GF^2\lt( -\s l^2+\g l\cot\xi \rt)D\varphi^1 \w
D\varphi^2+\frac{\g}{2}\tilde{d}_{\m\n}  dx^\m \w dx^\n.
\end{eqnarray}
This geometry is generally applicable when the axion field $\ta_1$
is pure gauge. For $\xi=\frac{\pi}{2}$ and $l=1$,
(\ref{eq:gen_grav_dual}) becomes the geometry obtained in the
appendix of \cite{Lunin:2005jy}.
%
\section{Discussion}\label{sec:discuss}
We have studied the gravity duals of the marginal deformation with a
complex parameter on two types of slanted $T^3$'s where the
direction along the third coordinate is not orthogonal to the
direction along the second coordinate or the direction along the
first coordinate. We have shown that in the supergravity solution
derived from eleven dimensional supergravity on the slanted
three-torus where the direction along the third coordinate is not
orthogonal to the direction along the second coordinate, the
intersection angle between them corresponds to the parameter of the
S-duality $SL(2,\mathbb{R})_s$ of the $\b$-deformation. In the
supergravity solution derived from eleven dimensional supergravity
on the slanted three-torus where the direction along the third
coordinate is not orthogonal to the direction along the first
coordinate, the intersection angle between them corresponds to the
parameter of the $SL(2,\mathbb{R})$ symmetry of type IIB
supergravity which is the symmetry of the toroidal compactification.
Therefore the intersection angle generalizes the $\b$-deformed
geometry. We have proposed a simpler form which is applicable when
the axion field is pure gauge.

\vspace{1cm}


\end{document}